\documentclass[doublecol]{epl2}

\usepackage{graphicx}

\begin{document}
\title{Structural properties of spatially embedded networks}

\author{Kosmas Kosmidis\inst {1}, Shlomo Havlin \inst {2}\and Armin Bunde\inst {1}} 
\institute{
	\inst {1} Institut f\"ur Theoretische Physik III, Justus-Liebig-Universit\"at Giessen, 35392 Giessen, Germany \\
	\inst {2} Minerva Center and Department of Physics, Bar-Ilan University, Ramat-Gan 52900, Israel
}

	{Complex systems}
	{Systems Obeying Scaling Laws}
 {Monte Carlo methods}

\date{\today}
\abstract{
We study the effects of spatial constraints on the structural properties of networks embedded in one or two dimensional space. When nodes are embedded in space, they have a well defined Euclidean distance $r$ between any pair. We assume that nodes at distance $r$ have a link with probability $p(r) \sim r^{- \delta}$.
We study the mean topological distance $l$ and the clustering coefficient $C$ of these networks and find that they both exhibit phase transitions for some critical value of the control parameter $\delta$ depending on the dimensionality $d$ of the embedding space. We have identified three regimes. When $\delta <d$, the networks are not affected at all by the spatial constraints. They are ``small-worlds'' $l\sim \log N$  with zero clustering at the thermodynamic limit. In the intermediate regime $d<\delta<2d$, the networks are affected by the space and the distance increases and becomes a power of $\log N$, and have non-zero clustering. When $\delta>2d$ the networks are ``large'' worlds $l \sim N^{1/d}$ with high clustering. Our results indicate that spatial constrains have a significant impact on the network properties, a fact that should be taken into account when modeling complex networks.}

\maketitle

\section{Introduction}
The interdisciplinary field of complex networks has recently received considerable attention \cite{DorMend,BarAlb,Strog1,AlbJeongBar,Vesp,Strog2,Watts,Newman,Havlin1,Havlin2,Gallos1}. A network is a set of nodes, representing the elementary units of a system. These nodes are connected by edges (bonds) when an interaction between the nodes is present. In a social network, for example, the nodes denote individuals. An edge connects two nodes if there is a relationship between the two individuals. Networks are particularly interesting as theoretical models of complex systems since they provide a straightforward way to describe all sorts of interactions between the elementary units of a system. 

Although there are infinitely many types of networks the attention of scientific community is actually focused in four types:
Regular networks, Random graphs, ``Small-world'' networks and scale-free networks. 

Regular Networks consist of nodes that are identical to each other.
The number of edges $k$ emanating from each node (the ``degree'' of the node) is a constant. 
Some of the most common constructions used in computational and theoretical physics, such as the square lattice, the Bethe lattice or the complete graph (a graph where each node is connected to every other node) belong to this type of networks. 

Random graphs were introduced and studied by Erd\"os and R\'enyi \cite{Erdos}. In Erd\"os -R\'enyi networks the distribution of the degrees of the nodes is Poissonian, and this fact facilitates the study of several important properties and allows analytical results. These networks have been extensively studied and used as some of the first models of social networks.

``Small-world'' networks were initially proposed by Watts and Strogatz as an improved model of social networks \cite{Strog2,Watts}.
 These networks show simultaneously the important feature of ``clustering'' (i.e. nodes that have common neighbors have a higher probability of being themselves connected) and the small-world phenomenon meaning that their topological diameter is slowly (logarithmically) increasing with increased system size. 

 In scale-free networks \cite{DorMend,BarAlb,AlbJeongBar} finally, the degree distribution follows a power law $P(k)\sim k ^{ -\gamma} $, with  $\gamma$ typically between two and three. Scale-free networks have the ultrasmall world property as the diameter of the network scales only logarithmically with the logarithm of the number of nodes \cite{Havlin1}.

In all the above cases, the spatial arrangement of nodes is considered of no importance. It is assumed that spatial constraints can be neglected such that mean field approaches become applicable. Many real networks are, however, embedded in 2D or 3D space. Examples are the Internet, airline networks, social networks, transportation and wireless communication networks \cite{DorMend,Greiner} etc. If these spatial constraints are important, mean field theory is no longer valid \cite{Stanley}, and new approaches have to be developed instead.
Spatially constrained networks are a promising type of complex networks whose importance has been only recently realized \cite{Sokolov,Rozenfeld,Tsallis,Mendes}. Up to now, most efforts deal with the simplest case, i.e. how to embed in the normal Euclidean space complex networks with a given degree distribution, while allowing only closest neighbor connections. If one allows additional long range connections then the resulting networks are more difficult to study but much closer to the real complex networks. 
 
Here we study  spatially constrained networks embedded in one and two dimensional space. We are interested in the simple case of networks that have a Poissonian degree distribution and where nodes are connected to each other with a probability $p(r) \sim r^{- \delta}$, i.e. depending on the Euclidean distance $r$ between the nodes. The choice of a power law for the distance distribution is supported from recent findings on the topology of real complex networks, such as the Internet \cite{BarPnas}, human travels \cite{Geisel} and in social networks \cite{Lil}.

 We propose a simple algorithm to generate such networks and we study properties such as the topological distance $l$ between nodes and the clustering coefficient $C$. The exponent $\delta$ has the role of a control parameter of the properties of the embedded networks and we find that there are three regimes of interest with qualitatively different behavior. Networks with $\delta < d$, where $d$ is the dimensionality of the embedding space, behave more or less like random graphs, networks with $\delta >2d$ behave much like regular lattices while networks with $\delta$ between these $d$ and $2d$ show intermediate power-law behavior with critical exponents that depend on $\delta$. 

Conceptually, our model is related to the ``small-world'' networks \cite{Strog2,Watts, Kleinberg} and to long-range percolation \cite{Benjamini, Coppersmith, Moukarzel}. In both the above cases, an underlying lattice structure is assumed and is ``perturbed'' by the addition of a fraction of long range connections. Thus, there is a competition between short and long range interactions which determines the properties of the network. In our model, no such underlying structure exists. A perfectly ordered lattice is recovered only in the limit of infinite $\delta$. There is, however, still a ``hidden'' competition between short and long range, but now the magnitude of both ``short'' and ``long'' is controlled by the exponent $\delta$ and the system size.

\section{Methods}   

To construct the networks, we first arrange the nodes in a one dimensional ($d=1$) linear chain or in a two dimensional ($d=2$) regular square lattice. Thus, between any pair of nodes there is a well defined Euclidean distance. Next, we initially choose a node $i$ randomly and assume that it is connected to $k_t$ other nodes. For each of these nodes, we first choose its distance $r$ from node $i$ with probability  $P(r) = c r^{d-1}r^{- \delta}$, where $c$ is determined from the normalization condition $\int_{1}^{L} P(r) dr =1$, with $L=N^{1/d}$. Then we pick one of the $N_r$ sites that are at distance $r$ within the underlying lattice. We repeat this process for all nodes $i$ in the underlying lattice and then remove multiple connections. The nodes of the final network we obtain this way do not have a fixed degree. Their mean degree $\bar{k}$ depends on both $k_t$ and $\delta$. In other words, for obtaining a certain $\bar{k}$ value for fixed $\delta$, we have to choose an appropriate value for $k_t$. When the networks are constructed we measure their actual degree probability distribution $p(k)$ to verify that all networks have at least approximately the same mean degree $\bar{k}$ and the actual distance distribution $p(r)$ i.e. the fraction of connected nodes that are at distance $r$ from each other, which is affected by the process of the removal of duplicate edges. In what follows we have succeeded in obtaining $\bar{k}\approx 4$ (to be precise $3.3 <\bar{k}\leq 4$) using $k_t=2$ when $\delta \leq 2.25$, $k_t=3$ when $2.25 < \delta \leq 3$ and $k_t=5$ for $\delta =4$ for the one dimensional case. For the two dimensional case $k_t=2$ produced networks with  $3.6 <\bar{k}\leq 4$ for all $\delta \leq 4.5$. Thus, no particular  ``fine-tuning'' was necessary.

\section{Quantities of interest}  
 We are interested in the mean topological distance $ l $ between the nodes of the network and the mean clustering coefficient $C$. The topological distance between two nodes $a$ and $b$ is the minimum number of links that a walker has to cross in order to arrive from $a$ to $b$. The (local) clustering coefficient of a node $i$ in a network is defined as:
\begin{equation} 
C(k_i) = \frac{t_i}{[k_i(k_i -1)/2]}
 \end{equation}
where $t_i$ is the number of triangles (loops of length 3) attached to this node divided by the maximum possible number of such loops.
For calculating the mean topological distance $l$ and the mean clustering coefficient $C$ we average over all nodes in the network (typically $10^5$) and over all network (typically $10^2$ ) realizations. 

\section{Results}
\begin{figure}
\begin{center}
\includegraphics[width=8.5cm, height=8.5cm]{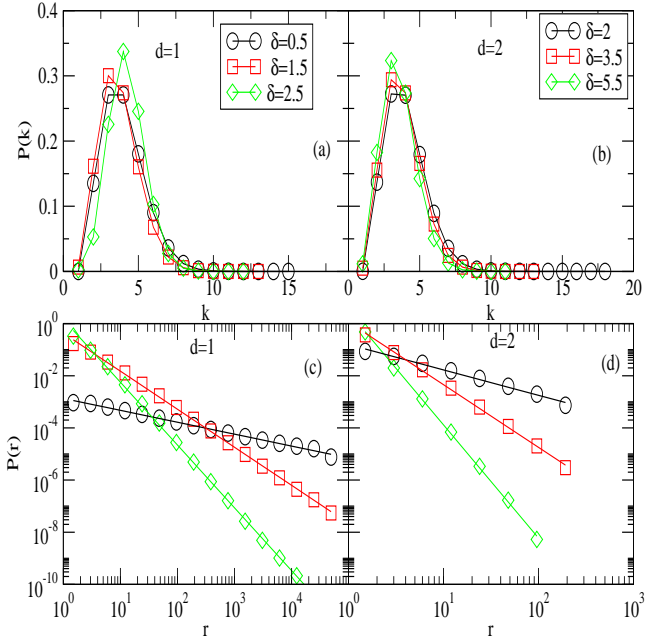}
\end{center}
\vspace{-0.8cm}
\caption{(a) Probability $P(k)$ that a node has degree $k$ for $d=1$ networks  with $N=10^5$,$\bar{k}\approx 4$ and for $\delta =0.5, 1.5$ and $2.5$. (b) Probability $P(k)$ for $d=2$ networks  with $N=5 \times 10^4$, $\bar{k} \approx 4$ and for $\delta =2,  3.5, 5.5$. (c) Probability $P(r)$ that a node has a connection at distance $r$ for the same $d=1$ networks as in (a). The points are logarithmically binned simulation data. (d) Probability $P(r)$ that a node has a connection at distance $r$ for the same $d=2$ networks as in (b). The points are logarithmically binned simulation data. In (c,d) the straight lines have slopes equal to $-(\delta -d+1)$.}
\label{fig1}
\end{figure} 

First, we show that the generated networks have the desired degree and distance distributions.
Figure \ref{fig1} shows the degree distribution $P(k)$ and the distance distribution $P(r)$ for the generated one dimensional (Fig \ref{fig1}(a,c)) and two dimensional (Fig \ref{fig1}(b,d)) networks, for three characteristic $\delta$ values each, $\delta=0.5,1.5,2.5$ ($d=1$) and  $\delta=2,3.5,5.5$ ($d=2$). Figures \ref{fig1}a,b confirm that the networks have a narrow degree distribution around $\bar{k}\approx 4$. Figures \ref{fig1}c,d confirm that $P(r)$ has the desired distance distribution $P(r) \sim r^{-(\delta -d +1)}$ for both $d=1$ and $d=2$.

\begin{figure}
\begin{center}
\includegraphics[width=8.5 cm,height=8.5cm,angle=0,clip]{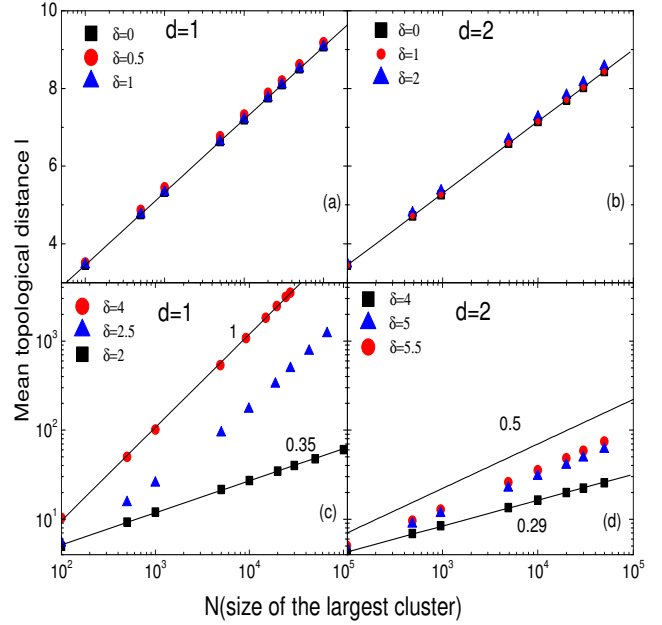}
\end{center}
\vspace{-0.8cm}
\caption{(a)Mean topological distance $l$ as a function of the networks size $N$ for $d=1$ networks  with $\bar{k}\approx 4$ and for $\delta =0, 0.5$ and $1$ i.e. in Regime 1 (LinearLog plot). (b)Mean topological distance $l$ as a function of the networks size $N$ for $d=2$ networks with $\bar{k} \approx 4$ and for $\delta =0, 1$ and $2$ i.e. in Regime 1 (LinearLog plot). The straight line is guide to the eye.  (c) Mean topological distance $l$ as a function of the networks size $N$ for $d=1$ networks with $\bar{k}\approx 4$ and for $\delta =2, 2.50$ and $4$ i.e. in Regime 3(LogLog plot).(d) Mean topological distance $l$ as a function of the networks size $N$ for $d=2$ networks with $\bar{k} \approx 4$ and for $\delta =4, 5, 5.5$ i.e. in Regime 3 (LogLog plot). The straight line has slope $0.5$. }
\label{fig2}
\end{figure}

Next, we study the mean topological distance $l$ as a function of the network size $N$. In the limiting case $\delta \rightarrow 0$ the spatial constraints are not significant and we expect that the networks will belong to the universality class of Erd\"os -R\'enyi random graphs, where $l$ scales as $l \sim \log N$. In the opposite case $\delta \rightarrow \infty$, the spatial constraints will result in an ordered lattice-like structure that belongs to the universality class of regular lattices, where the mean topological distance $l$ scales as $N^{1/d}$ ( $d$ is the dimensionality of the embedding space). We are interested in how the dependence of the topological distance $l$ on $N$ changes when $\delta$ changes from zero to infinity.

\begin{figure} 
\begin{center}
\includegraphics[width=8cm,height=8cm]{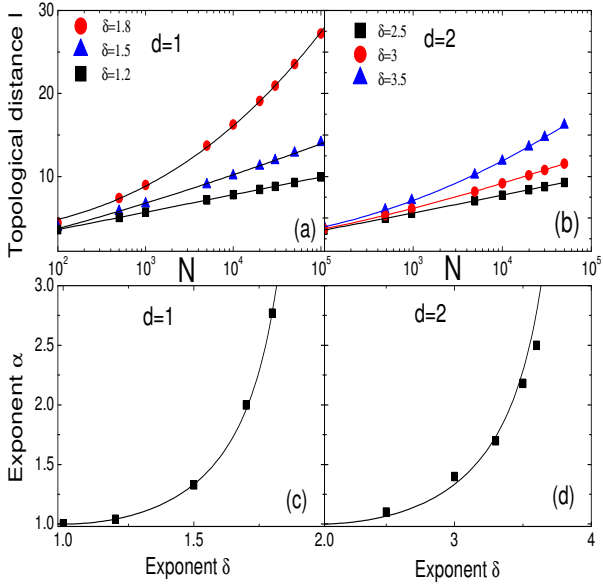}
\end{center}
\vspace{-0.8cm}
\caption{(a) Mean topological distance $l$ as a function of the networks size $N$ for $d=1$ networks  with $\bar{k}\approx 4$ and for $\delta =1.2, 1.5$ and $1.8$ (bottom to top) i.e. in Regime 2 (LinearLog plot). The lines are fits to the simulation data of the scaling assumption (Eq. \ref{eqlB}).(b) Mean topological distance $l$ as a function of the networks size $N$ for $d=2$ networks , $\bar{k} \approx 4$ and for $\delta =2.5,3,3.5$ (bottom to top) i.e. in Regime 2 (LinearLog plot). Points are simulation data and lines are fits of Eq. \ref{eqlB}.(c) Exponent $\alpha$ as a function of $\delta$ for $d=1$ networks in the second regime ($1 < \delta <2$). (d) Exponent $\alpha$ as a function of $\delta$ for $d=2$ networks in the second regime ($2 < \delta <4$). Points are best fit estimates to the simulation data. Lines are plots of Eq.(\ref{expa})}
\label{fig3}
\end{figure}

Our results (Figs. \ref{fig2} and \ref{fig3}) indicate that in both $d=1$ and $d=2$, there are three regimes separated by $\delta=d$ and $\delta=2d$, where the behavior of the networks is qualitatively different. In the first regime $0 \leq \delta <d$ (see Fig. \ref{fig2}a for $d=1$ and Fig. \ref{fig2}b for $d=2$), $l$ scales as $\log N$ independent of $\delta$. In the third regime, $\delta > 2d$ 
(see Fig. \ref{fig2}c for $d=1$ and Fig. \ref{fig2}d for $d=2$) the result suggests that $l$ asymptotically scales as $N^{1/d}$ as for regular lattices. This dependence is well observed in $d=1$ (Fig. \ref{fig2}c) for $\delta =4$ and in $d=2$ for $\delta=5.5$. For $\delta$ closer to $2d$, this scaling is only recovered for sufficiently large networks. For $\delta = 2d$, $l$ scales as $N^\gamma$, with $\gamma \approx 0.35$ for $d=1$ and $\gamma \approx 0.3$ for $d=2$ (see the lowest curves in Figs. \ref{fig2}c and d, respectively).

Figure \ref{fig3}(a,b)  shows the mean topological distance $l$ as a function of $N$ in the intermediate regime $d<\delta<2d$.
The plots suggest that for both $d=1$ and $d=2$, $l$ scales as  $l \sim (\ln(N))^\alpha$. 
Accordingly, we can summarize the dependence of $l$ on $N$ as follows:

 \begin{equation} 
l \sim \left\{
             \begin{array}{ll} 
                 			\log(N) & \delta \leq d  \\
               				(\log(N))^\alpha & d \leq \delta <2d   \\
					N^{1/d} &  \delta >2d         
               \end{array}      
                \right.
\label{eqlB}
 \end{equation} 
When $\delta=2d$ the numerical results are quite well described by a power law $N^{\gamma}$ where  $\gamma \approx 0.35$ for $d=1$ and $\gamma \approx 0.3$ for $d=2$. It is, however, probable that there are logarithmic corrections to this scaling form.
The variation of the exponent $\alpha$ with $\delta$ is shown in Fig. \ref{fig3}c for $d=1$ and  Fig. \ref{fig3}d for $d=2$. The results suggest that approximately

\begin {equation} 
\alpha =\left\{
	\begin{array}{ll} 
		\frac{1/\delta}{2-\delta}  & d=1 \\
 		\frac{4/\delta}{4-\delta}  & d=2 \\
	\end{array}      
         \right.
\label{expa}
 \end{equation}  
Equation (\ref{expa}) provides a good fit to the results in Figs. \ref{fig3}(c,d), and although is not a rigorous analytic result it may be useful for practical purposes \footnote{Of course, in Eq. (\ref{eqlB}) in the range $d \leq \delta <2d$  a more complex scaling form cannot be excluded.}
Note that Eq. (\ref{expa}) ensures a smooth transition between the first and second regime. To match with the power law behavior when $\delta>2d$, $\alpha$ diverges when approaching $\delta=2d$ from below. In $d=2$, also a weaker singularity could fit the data.

We like to note that these transition points arise naturally in the scaling of the mean distance $\bar{r}$ and the maximum distance  $r_{max}$ with the system size $L$. By definition $\bar{r}=\int_1^L r p(r)dr$, while $r_{max}$ is obtained from the condition $L^d \int_{r_{max}}^L p(r) dr \approx 1$. It is easy to verify that

\begin {equation} 
\bar{r} \sim \left\{
		\begin{array}{ll} 
		L	  & \delta \leq d \\
 		L^{d-\delta +1} +c_1 & \delta > d \\
	\end{array}      
         \right.
\label{eqmeanr}
 \end{equation} 
where $c_1$ does not depend on $N$ and
\begin {equation} 
r_{max} \sim \left\{
		\begin{array}{ll} 
		L	  & 0 < \delta \leq 2d \\
 		L^{\frac{d}{\delta-d}} & \delta \geq 2d \\
	\end{array}      
         \right.
\label{eqmaxr}
 \end{equation} 
Accordingly, the relative mean distance $\bar{r}/L$ is finite for $\delta \leq d$ and 0 for $\delta > d$ in the limit of large $L$. Similarly, the relative maximum distance  $r_{max}/L$ is finite for $\delta \leq 2d$ and 0 for $\delta > 2d$ for $L$ going to infinity. The above relations can qualitatively elucidate the existence of different regimes. For simplicity consider the case $d=1$ and assume one attempts to go from the one end of the ``chain'' to the other. When the mean length $\bar{r}$ of each link is comparable to the lattice size $L$ then he can succeed with only one step on average. The lattice is a small world (regime 1). When the the mean length $\bar{r}$ is less than $L$ but $r_{max}$ is of the order of $L$, then the walker will have to do more steps but he will eventually arrive at a site that has a long edge $r_{max}$ that will allow him to arrive at the end of the lattice. The lattice is not a small world as in the first case but also not a regular lattice since there are links that decrease the topological distance considerably (regime 2). When both $\bar{r}$ and $r_{max}$ are less than $L$, there are no shortcuts. Renormalizing the graph will result in a regular linear chain (regime 3).

\begin{figure} 
\begin{center}
\includegraphics[width=8.5cm,height=8.5cm]{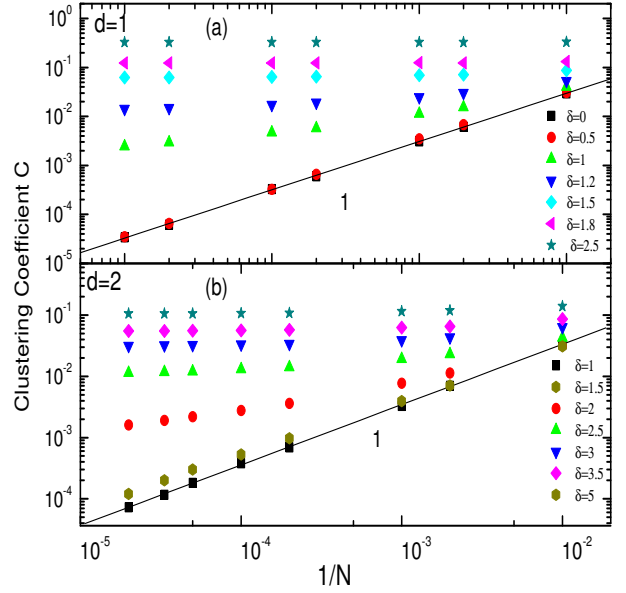}
\end{center}
\vspace{-0.8cm}
\caption{(a) Clustering coefficient $C$ as a function of $1/N$ for $\delta=0,0.5,1,1.2,15,1.8,2.5$ and $d=1$. (b) Clustering coefficient $C$ as a function of $1/N$ for $\delta=1,1.5,2,2.5,3,3.5,5$ and $d=2$. Straight lines are best fits with slope 1.}
\label{fig4}
\end{figure}

Finally, we study how the clustering coefficient $C$ depends on $\delta$ and $N$ (Fig.\ref{fig4}). We are particularly interested in the behavior at the transition points $\delta=d$ and $\delta =2d$. We expect that for $\delta \rightarrow 0$, $C \sim 1/N$ similar to the case of Erd\"os -R\'enyi random graphs. When  $\delta \rightarrow \infty$ we expect the (high) clustering of the lattices. We find that the critical point at $\delta=d$ separates a phase of zero clustering when $\delta< d$ from a phase of finite clustering when $\delta>d$. The second transition point at $\delta=2d$ does not influence $C$ remarkably. Figures \ref{fig4}(a,b) show $C$ as a function of $1/N$ for $d=1$ and $d=2$ respectively. The clustering coefficient $C$ decreases as $1/N$ for $\delta <d$ in both one and two dimensions, while for $\delta> d$ the slope of the curves in the double logarithmic plot becomes zero, confirming the existence of a regime with non zero clustering. It is remarkable that as soon as $\delta$ becomes larger than $d$, the spatial effects become very pronounced. The topology not only affects the mean topological distance but also increases simultaneously the clustering coefficient.

\section{Conclusions}
We have examined the mean topological distance $l$ and the clustering coefficient $C$ of networks that are embedded in one or two dimensional space. 
Both exhibit phase transitions for some critical value of the control parameter $\delta$. We have identified three regimes. When $\delta <d$, where $d$ is the dimensionality of the embedding space, the networks are ``small-worlds'' with zero clustering at the thermodynamic limit. In the intermediate regime $d<\delta<2d$, the topology changes. The topological distance becomes a power of the logarithm of the system size and the networks have non-zero clustering. When $\delta>2d$ the networks do not have the small world property and have high clustering. Our results indicate that spatial constraints have an important influence on the network properties and this fact should be taken into account when modeling real complex networks. A natural extension of this work will be to consider fractal or multifractal substrates instead of regular lattices \cite{Greiner, benAv} and study their influense on the topological properties of the embeded networks.

{\it Acknowledgment:}  This work was supported by a European research NEST Project No DYSONET 012911.

\end{document}